%% file: ms.tex
\documentclass[sigconf]{acmart}

\usepackage{booktabs}
\usepackage{subcaption}

\usepackage[utf8]{inputenc}
\usepackage[greek,english]{babel}

\usepackage{proof}
\usepackage{amsmath}
\usepackage{amssymb}
\usepackage{amsthm}
\usepackage{syntax}
\usepackage{float}

\input{slti-macros}

\acmConference[IFL'18]{International Symposium on Implementation and
  Application of Functional Languages}{August 2019}{Lowell, MA, USA}
\acmYear{2019}
\copyrightyear{2019}
\setcopyright{none}
\acmDOI{}
\acmISBN{}

\begin{document}

\input{slti-title}

\maketitle

\input{slti-intro}

\input{slti-rules}
\input{slti-discussion}

\bibliographystyle{ACM-Reference-Format}
\bibliography{slti-bib}

\end{document}

%% file: slti-macros.tex
\DeclareUnicodeCharacter{03BB}{\ensuremath{\lambda}}
\DeclareUnicodeCharacter{039B}{\ensuremath{\Lambda}}
\DeclareUnicodeCharacter{2200}{\ensuremath{\forall}}
\DeclareUnicodeCharacter{2200}{\ensuremath{\forall}}
\DeclareUnicodeCharacter{2115}{\ensuremath{\mathbb{N}}}
\DeclareUnicodeCharacter{1D539}{\ensuremath{\mathbb{B}}}
\DeclareUnicodeCharacter{2192}{\ensuremath{\to}}
\DeclareUnicodeCharacter{00D7}{\ensuremath{\times}}
\DeclareUnicodeCharacter{27E8}{\ensuremath{\langle}}
\DeclareUnicodeCharacter{27E9}{\ensuremath{\rangle}}

\newcommand{\refsec}[1]{Section \ref{sec:#1}}
\newcommand{\labsec}[1]{\label{sec:#1}}
\newcommand{\reffig}[1]{Figure \ref{fig:#1}}
\newcommand{\labfig}[1]{\label{fig:#1}}
\newcommand{\refthm}[1]{Theorem \ref{thm:#1}}
\newcommand{\labthm}[1]{\label{thm:#1}}

\newcommand{\elab}{\rightsquigarrow}
\newcommand{\vars}[1]{{\overline{#1}}}
\newcommand{\lowerdec}[1]{\lfloor {#1} \rfloor}
\newcommand{\subid}{\sigma_{id}}
\newcommand{\To}{\Rightarrow}
\newcommand{\nada}{\varnothing}
\newcommand{\parr}{?\!\to }

\newcommand{\df}{\vdash}
\newcommand{\decdir}{\vdash_{\delta}}
\newcommand{\decsyn}{\vdash_{\Uparrow}}
\newcommand{\decchk}{\vdash_{\Downarrow}}
\newcommand{\shim}{\vdash^{\text{I}}}
\newcommand{\meta}{\vdash^{\text{P}}}
\newcommand{\metaapp}{\vdash^{\cdot}}

\newcommand{\algdir}{\Vdash_{\delta}}
\newcommand{\algsyn}{\Vdash_{\Uparrow}}
\newcommand{\algchk}{\Vdash_{\Downarrow}}
\newcommand{\algapp}{\Vdash^{\cdot}}

\newcommand{\proto}{\Vdash^{\text{?}}}
\newcommand{\match}{\Vdash^\text{:=}}

\newcommand{\ann}[2]{#1\! : \! #2}
\newcommand{\dec}[2]{#1\!=\!#2}
\newcommand{\aabs}[4]{{#1}\, \ann{#2}{#3}.\, #4}
\newcommand{\dabs}[3]{\forall\, \dec{#1}{#2}.\, #3}
\newcommand{\abs}[3]{{#1}\, #2.\, #3}

\newcommand{\introx}[2]{#1 \to #2 \to ⟨#1 \times #2⟩}
\newcommand{\lamid}{\abs{\lambda}{x}{x}}
\newcommand{\lamnatid}{\aabs{\lambda}{x}{ℕ}{x}}
\newcommand{\pair}[2]{⟨#1\times#2⟩}
\newtheoremstyle{case}{}{}{}{}{}{:}{ }{}
\theoremstyle{case}

%% file: slti-title.tex
\title{Spine-local Type Inference}

\author{Christopher Jenkins}
\affiliation{%
  \institution{University of Iowa}
}
\email{christopher-jenkins@uiowa.edu}

\author{Aaron Stump}
\affiliation{%
  \institution{University of Iowa}
}
\email{aaron-stump@uiowa.edu}

\begin{abstract}
  We present \textit{spine-local} type inference, a partial type
inference system for inferring omitted type annotations for System F
terms based on local type inference. \textit{Local type inference}
relies on bidirectional inference rules to propagate type information
into and out of adjacent nodes of the AST and restricts type-argument
inference to occur only within a single node. Spine-local inference
relaxes the restriction on type-argument inference by allowing it to
occur only within an \textit{application spine} and improves upon it
by using \textit{contextual type-argument inference}. As our goal is
to explore the design space of local type inference, we show that,
relative to other variants, spine-local type inference enables desirable
features such as first-class curried applications, partial type
applications, and the ability to infer types for some terms not
otherwise possible. Our approach enjoys usual properties of a
bidirectional system of having a specification for our inference
algorithm and predictable requirements for typing annotations, and in
particular maintains some the advantages of local type inference such
as a relatively simple implementation and a tendency to produce
good-quality error messages when type inference fails.
\end{abstract}

\ccsdesc[500]{Software and its engineering~Language features}

\keywords{bidirectional typechecking, polymorphism, type errors}

%% file: slti-intro.tex
\section{Introduction}

Local type inference\cite{PT98} is a simple yet effective partial
technique for inferring types for programs. In contrast to complete
methods of type inference such as the Damas-Milner system\cite{DM82}
which can type programs without any type annotations by restricting
the language of types, \textit{partial} methods require the programmer
to provide some type annotations and, in exchange, are suitable for
use in programming languages with rich type features such as
impredicativity and subtyping\cite{PT98,OZZ01}, dependent
types\cite{XP99}, and higher-rank types\cite{PVWS07}, where complete
type inference may be undecidable.

Local type inference is also contrasted with \textit{global} inference
methods (usually based on unification) which are able to infer more
missing annotations by solving typing constraints generated from the
entire program. Though more powerful, global inference methods can
also be more difficult for programmers to use when type inference
fails, as they can generate type errors whose root cause is distant
from the location the error is reported\cite{Ma02}. Local type
inference address this issue by only propagating typing information between
adjacent nodes of the abstract syntax tree (AST), allowing programmers
to reason \textit{locally} about type errors. It achieves this by
using two main techniques: \textit{bidirectional type inference rules}
and \textit{local type-argument inference}.

The first of these techniques, bidirectional type inference, is not
unique to local type inference (\cite{DK13,PVWS07,VWP06,XO18} are just
a few examples), and uses two main judgment forms, often called
\textit{synthesis} and \textit{checking} mode. When a term $t$
synthesizes type $T$, we view this typing information as coming up and
out of $t$ and as available for use in typing nearby terms; when $t$
checks against type $T$ (called in this paper the \textit{contextual}
type), this information is being pushed down and in to $t$ and is
provided by nearby terms.

The second of these techniques, local type-argument inference, finds
the missing types arguments in polymorphic function applications by
using only the type information available at an application node of
the AST. For a simple example, consider the expression $\texttt{id}\
z$ where \texttt{id} has type $\abs{\forall}{X}{X \to X}$ and $z$ has
type ℕ. Here we can perform \textit{synthetic} type-argument inference
by synthesizing the type of $z$ and comparing this to the type of
\texttt{pair} to infer that the missing type argument instantiating $X$ is ℕ.

Using these two techniques, local type inference has a number of
desirable properties. Though some annotations are still required, in
practice a good number of type annotations can be omitted, and often
those that need to remain are predictable and coincide with
programmers' expectations that they serve as useful and
machine-checked documentation\cite{PT98,HP99}. Without further
instrumentation, local type inference already tends to report type
errors close to where further annotations are required; more recently,
it has been used in \cite{Pl16} as the basis for developing autonomous
type-driven debugging and error explanations. The type inference
algorithms of \cite{PT98,OZZ01} admit a specification for their
behavior, helping programmers understand why certain types were
inferred without requiring they know every detail of the
type-checker's implementation. Add to this its relative simplicity and
robustness when extended to richer type systems and it seems
unsurprising that it has been a popular choice for type inference in
programming languages.

Unfortunately, local type inference can fail even when it seems like
there should be enough typing information available locally. Consider
trying to check that the expression $\texttt{pair}\ (\lamid)\ z$ has
type $⟨(ℕ \to ℕ) \times ℕ⟩$, assuming $\texttt{pair}$ has type
$\abs{\forall}{X}{\abs{\forall}{Y}{X \to Y \to ⟨X \times Y⟩}}$).
The inference systems presented in \cite{OZZ01,PT98} will fail here
because the argument $\lamid$ does not synthesize a type.
The techniques proposed in the literature of local type inference for
dealing with cases similar to this include classifying and
avoiding such ``hard-to-synthesize'' terms\cite{HP99} and
utilizing the partial type information provided by polymorphic
functions\cite{OZZ01}; the former was dismissed as unsatisfactory by
the same authors that introduced it and the latter is of no help 
in this situation, since the type of \texttt{pair} tells us nothing
about the expected type of $\lamid$. What we need in this case is
\textit{contextual} type-argument inference, utilizing the information
available from the expected type of the whole application to know
argument $\lamid$ is expected to have type $ℕ \to ℕ$.

Additionally, languages using local type inference usually use
fully-uncurried applications in order to maximize the notion of
``locality'' for type-argument inference, improving its
effectiveness. The programmer can still use curried applications if
desired, but ``they are second-class in this respect.''\cite{PT98}. It
is also usual for type arguments to be given in an ``all or nothing''
fashion in such languages, meaning that even if only \textit{one} cannot be
inferred, all must be provided. We believe that currying and partial
type applications are useful idioms for functional programming and
wish to preserve them as first-class language features.

\subsection{Contributions}

In this paper, we explore the design space of local type inference in
the setting of System F\cite{Gi86,GTL89} by developing
\textit{spine-local} type inference, an approach that both expands the
locality of type-argument inference to an \textit{application spine}
and augments its effectiveness by using the \textit{contextual} type
of the spine. In doing so, we
\begin{itemize}
\item show that we can restore first-class currying, partial type
  applications, and infer the types for some
  ``hard-to-synthesize'' terms not possible in other variants of local
  type inference;
\item provide a specification for contextual
  type-argument inference with respect to which we show our algorithm
  is sound and complete
\item give a weak completeness theorem for our type system with
  respect to fully annotated System F programs, indicating the
  conditions under which the programmer can expect type inference
  succeeds and where additional annotations are required when it fails.
\end{itemize}
\noindent Spine-local type inference is being implemented in
Cedille\cite{St17}, a functional programming language with
higher-order and impredicative polymorphism and dependent types and
intersections. Though the setting for this paper is much simpler, we
are optimistic that spine-local type inference will serve as a good
foundation for type inference in Cedille that makes using its rich
type features more convenient for programmers.

The rest of this paper is organized as follows: in \refsec{syntax} we
cover the syntax and some useful terminology for our setting; in
\refsec{drules} we present the type inference rules constituting a
specification for contextual type-argument inference, consider its
annotation requirements, and illustrate its use, limitations, and the
type errors it presents to users; in \refsec{arules} we show
the prototype-matching algorithm implementing contextual type-argument
inference; and in \refsec{discussion} we discuss how this work
compares to other approaches to type inference.

%% file: slti-rules.tex
\section{Internal and External Language}
\labsec{syntax}
Type inference can be viewed as a relation between an
\textit{internal} language of terms, where all needed typing
information is present, and an \textit{external} language, in which
programmers work directly and where some of this information can be
omitted for their convenience. Under this view, type inference for the
external language not only associates a term with some type but also with
some \textit{elaborated} term in the internal language in which all
missing type information has been restored. In this section, we
present the syntax for our internal and external languages as well as
introduce some terminology that will be used throughout the rest of
this paper.

\subsection{Syntax}
We take as our internal language explicitly typed System F (see
\cite{GTL89}); we review its syntax below:
\begin{align*}
  \textbf{Types}    && S,T,U,V ::=\ & X,Y,Z\ |\ S \to T\ |\ \abs{\forall}{X}{T} \\
  \textbf{Contexts} && \Gamma ::=\ & \cdot\ |\ \Gamma,X\ |\
                                     \Gamma,\ann{x}{T} \\
  \textbf{Terms}    && e,p ::=\ & x\ |\ \aabs{\lambda}{x}{T}{e}\
                        |\ \abs{\Lambda}{X}{e}\ |\ e\ e'\ |\ e [T]
\end{align*}
Types consist of type variables, arrow types, and type quantification,
and typing contexts consist of the empty context, type variables (also
called the context's \textit{declared} type variables), and term
variables associated with their types. The internal language of terms
consists of variables, λ-abstractions with annotations on bound
variables, Λ-abstractions for polymorphic terms, and term and type
applications. Our notational convention in this paper is that
term meta-variable $e$ indicates an elaborated term for which
all type arguments are known, and $p$ indicates a
\textit{partially} elaborated term where some type
arguments are type meta-variables (discussed in \refsec{drules}).

The external language contains the same terms as the internal language
as well as bare λ-abstractions -- that is, λ-abstractions missing an
annotation on their bound variable:
\begin{align*}
  \textbf{Terms} & & t,t' ::= x\ |\ \aabs{\lambda}{x}{T}{t}\
                     |\ \abs{\lambda}{x}{t}\ |\ \abs{\Lambda}{X}{t}\
                     |\ t\ t'\ |\ t [T]
\end{align*}
 Types and contexts are the same as for the internal language and are omitted.
\subsection{Terminology}
In both the internal and external languages, we say that the
\textit{applicand} of a term or type application is the term in the
function position. A \textit{head} a is either a variable or
λ-abstraction (bare or annotated), and an \textit{application
spine}\cite{CP03} (or just \textit{spine}) is a view of an application
as consisting of some head (called the \textit{spine head}) followed
by a sequence of (term and type) arguments. The \textit{maximal
application} of a sub-expression is the spine in which it occurs as an
applicand, or just the sub-expression itself if it does not. For example,
spine $x[S]\ y\ z$ is the maximal application of itself and its
applicand sub-expressions $x$, $x[S]$, and $x[S]\ y$, with $x$ as head
of the spine. Predicate $App(t)$ indicates term $t$ is some term or
type application (in either language) and we define it formally as
${(\exists\ t_1,t_2.\ t = t_1\ t_2)}\ \lor\ {(\exists\ t',S.\ t =
t'[S])}$.

Turning to definitions for types and contexts, function $DTV(\Gamma)$
calculates the set of \textit{declared type variables} of context $\Gamma$
and is defined recursively by the following set of equations:
\begin{eqnarray*}
     DTV(\cdot) & =  & \nada
  \\ DTV(\Gamma,X) & = & DTV(\Gamma) \cup \{X\}
  \\ DTV(\Gamma,\ann{x}{T}) & = & DTV(\Gamma)
\end{eqnarray*}
\noindent Predicate $WF(\Gamma,T)$ indicates that type $T$ is
\textit{well-formed} under $\Gamma$ -- that is, all free type variables of
$T$ occur as declared type variables in $\Gamma$ (formally $FV(T) \subseteq
DTV(\Gamma)$).

\section{Type Inference Specification}
\labsec{drules}
\begin{figure*}
  \centering
  \[
    \begin{array}{ccc}
      \fbox{$\Gamma \decdir t : T \elab e$}
      & \infer[Var]{\Gamma \decdir x:\Gamma(x) \elab x}{}
      & \infer[Abs]
        {\Gamma \decchk \abs{\lambda}{x}{t} : T \to S \elab \aabs{\lambda}{x}{T}{e}}
        {\Gamma,\ann{x}{T} \decchk t : S \elab e}
      \\ \\ \infer[AAbs]
        {\Gamma \decdir \aabs{\lambda}{x}{T}{t} : T \to S \elab \aabs{\lambda}{x}{T}{e}}
        {\Gamma,\ann{x}{T} \decdir t : S \elab e}
      & \infer[TAbs]
      {\Gamma \decdir \abs{\Lambda}{X}{t} : \forall X . T \elab \abs{\Lambda}{X}{e}}
      {\Gamma,X \decdir t : T \elab e}
      & \infer[TApp]
        { \Gamma \decdir t [S] : [S/X]T \elab e [S] }
        { \Gamma \decsyn t : \forall X . T \elab e}
    \end{array}
  \] \\ \[
    \begin{array}{cc}
      \infer[AppSyn]
      { \Gamma \decsyn t\ t' : T \elab e}
      { \Gamma ; ? \shim t\ t' : T \elab (e,\subid)
        & MV(\Gamma, e) = \varnothing }
      & \infer[AppChk]
        { \Gamma \decchk t\ t' : \sigma\ T \elab \sigma\ p}
        { \Gamma ; \sigma\ T \shim t\ t' : T \elab (p,\sigma)
          & MV(\Gamma, p) = dom(\sigma)
        }
    \end{array}
  \]
  \caption{Bidirectional inference rules with elaboration}
  \labfig{bidir}
\end{figure*}
\begin{figure*}
  \begin{subfigure}{1\linewidth}
    \caption{Shim (specification)}
    \labfig{decshim}
    \centering
    \[
      \begin{array}{cc}
        T_? ::= T\ |\ ?
        & \infer
          { \Gamma ; T_? \shim t\ t' : T \elab (p, \sigma)
          }
          { \Gamma \meta t\ t' : T \elab (p, \sigma)
            & MV(\Gamma,T) = dom(\sigma)
          }
      \end{array}
    \]
\end{subfigure}
\begin{subfigure}{1\linewidth}
  \caption{\fbox{$\Gamma \meta t : T \elab (p, \sigma)$}}
  \centering
  \[
    \begin{array}{ccc}
      \infer[PHead]
      { \Gamma \meta t : T \elab (e, \subid) }
      { \neg App(t)
         & \Gamma \decsyn t : T \elab e
      }
      & \infer[PTApp]
      { \Gamma \meta t[S] : [S/X] T \elab (p[S], \sigma) }
      { \Gamma \meta t : \forall X . T \elab (p, \sigma) }
      & \infer[PApp]
      { \Gamma \meta t\ t' : T' \elab (p', \sigma') }
      { \Gamma \meta t : T \elab (p, \sigma)
        & \Gamma \metaapp (\ann{p}{T}, \sigma) \cdot t' : T' \elab
          (p', \sigma')}
    \end{array}
  \]
  \labfig{meta}
\end{subfigure}
\begin{subfigure}{1\linewidth}
  \centering
  \caption{\fbox{$\Gamma \metaapp (\ann{p}{T},\sigma) \cdot t' : T' \elab (p',
      \sigma')$}}
  \[
    \begin{array}{cc}
      \infer[PForall]
      { \Gamma \metaapp (\ann{p}{\abs{\forall}{X}{T}},\sigma) \cdot t' : T' \elab
      (p', \sigma') }
      { \sigma'' \in \{\sigma,[S/X] \circ \sigma\}
        & WF(\Gamma,S)
        & \Gamma \metaapp (\ann{p[X]}{T},\sigma'') \cdot t' : T' \elab
          (p', \sigma')
      }
      &
      \infer[PChk]
      { \Gamma \metaapp (\ann{p}{S \to T}, \sigma) \cdot t' : T \elab (p\ e', \sigma)}
      { MV(\Gamma,\sigma\ S) = \varnothing
      & \Gamma \decchk t' : \sigma\ S \elab e'
      &  }
      \\ \\ \infer[PSyn]
        { \Gamma \metaapp (\ann{p}{S \to T},\sigma) \cdot t' :
      \vars{[U/Y]}\ T \elab ((\vars{[U/Y]}\ p)\ e', \sigma)}
        { MV(\Gamma,\sigma\ S) = \vars{Y} \neq \varnothing
          & \Gamma \decsyn t' : \vars{[U/Y]}\ \sigma\ S \elab e'
        }
    \end{array}
  \]
  \labfig{metaapp}
\end{subfigure}
\caption{Specification for contextual type-argument inference}
\labfig{dec}
\end{figure*}

The typing rules for our internal language are standard for explicitly
typed System F and are omitted (see Ch. 23 of \cite{TAPL02} for a
thorough discussion of these rules). We write ${\Gamma \df e : T}$ to
indicate that under context $\Gamma$ internal term $e$ has type
$T$. For type inference in the external language, \reffig{bidir} shows
judgment $\decdir$ which consists mostly of standard (except for
$AppSyn$ and $AppChk$) bidirectional inference rules with elaboration
to the internal language, and \reffig{dec} shows the specification for
contextual type-argument inference. Judgment $\meta$ in \reffig{meta}
handles traversing the spine and judgment $\metaapp$ in
\reffig{metaapp} types its term applications and performs
type-argument inference (both synthetic and
contextual). \reffig{decshim} gives a ``shim'' judgment $\shim$ which
bridges the bidirectional rules with the specification for rhetorical
purposes (discussed below). Though these rules are not algorithmic,
they \textit{are} syntax-directed, meaning that for each judgment the
shape of the term we are typing (i.e. the \textit{subject} of typing)
uniquely determines the rules that applies.

\paragraph{Bidirectional Rules}
We now consider more closely each judgment form and its rules starting
with $\decdir$, the point of entry for type inference. The two
modes for type inference, checking and synthesizing, are
indicated resp. by $\decchk$ (suggesting pushing a type down and
into a term) and $\decsyn$ (suggesting pulling a type up and out
of a term). Following the notational convention of Peyton Jones et
al.\cite{PVWS07} we abbreviate two inference rules that differ only in
their direction to one by writing $\decdir$, where $\delta$ is a
parameter ranging over $\{\Uparrow,\Downarrow\}$. We read judgment
${\Gamma \decsyn t : T \elab e}$ as: ``under context $\Gamma$, term
$t$ synthesizes type $T$ and elaborates to $e$,'' and a similar
reading for checking mode applies for $\decchk$. When the direction
does not matter, we will simply say that we can \textit{infer} $t$ has
type $T$.

Rule $Var$ is standard. Rule $Abs$ says we can infer missing type
annotation $T$ on a λ-abstraction when we have a contextual arrow type
$T \to S$. Rules $AAbs$ and $TAbs$ say that Λ- and annotated
λ-abstractions can have their types either checked or
synthesized. $TApp$ says that a type application $t[S]$ has its type
inferred in either mode when the applicand $t$ synthesizes a
quantified type. The reason for this asymmetry between the modes of the
conclusion and the premise is that even when in checking mode,
it is not clear how to work backwards from type $[S/X]T$ to
$\abs{\forall}{X}{T}$.

$AppSyn$ and $AppChk$ are invoked on maximal applications and are the
first non-standard rules. To understand how these rules work, we must
1) explain the ``shim'' judgment $\shim$ serving as the interface for
spine-local type-argument inference and 2) define meta-language
function $MV$. Read $\Gamma ; T_? \shim t\ t' : T \elab (p,\sigma)$
as: ``under context $\Gamma$ and with (optional) contextual type
$T_?$, partially infer application $t\ t'$ has type $T$ with
elaboration $p$ and solution $\sigma$,'' where $\sigma$ is a
substitution mapping a some meta-variables (i.e. omitted type
arguments) in $p$ to contextually-inferred type arguments.

In rule $AppSyn$, $?$ is provided to $\shim$ indicating no contextual
type is available. We constrain $\sigma$ to be the identity
substitution (written $\subid$) and that elaborated term $p$ has no
unsolved meta-variables, matching our intuition that all type
arguments must be inferred synthetically. In rule $AppChk$, we provide
the contextual type to $\shim$ and check (implicitly) that it equals
$\sigma\ T$ and (explicitly) that all remaining meta-variables in $p$
are solved by $\sigma$, then elaborate $\sigma\ p$ (the replacement of
each meta-variable in $p$ with its entry in $\sigma$). Shared by both
is the second premise of the (anonymous) rule introducing $\shim$ that
$\sigma$ solves precisely the meta-variables of the partially inferred
type $T$ for application $t\ t'$.

\paragraph{Meta-variables}
What are the ``meta-variables'' of elaborations and types? When $t$ is
a term application with some type arguments omitted in its spine, its
partial elaboration $p$ from spine-local type-argument inference under
context $\Gamma$ fills in each missing type argument with either a
well-formed type or with a \textit{meta-variable} (a type variable not
declared in $\Gamma$) depending on whether it was inferred
synthetically. For example, if $t = \texttt{pair}\
(\abs{\lambda}{x}{x})\ z$ and we wanted to check that it has type $T =
⟨(ℕ → ℕ) \times ℕ⟩$ under a typing context $\Gamma$ associating
\texttt{pair} with type $\abs{\forall}{X}{\abs{\forall}{Y}{X \to Y \to
⟨X \times Y⟩}}$ and $z$ with type ℕ, then we could derive
\begin{align*}
\Gamma ; T \shim t : ⟨X \times ℕ⟩ \elab (\texttt{pair}[X][ℕ]\
  (\aabs{\lambda}{x}{ℕ}{x})\ z, [ℕ → ℕ/X])
\end{align*}
\noindent (assuming some base type ℕ, some family of base types $⟨S
\times T⟩$ for all types $S$ and $T$, and assuming $X$ is not declared
in $\Gamma$.) Looking at the partial elaboration of $t$, we would see
that type argument $X$ was inferred from its contextual type $⟨(ℕ → ℕ)
\times ℕ⟩$ and that $Y$ was inferred from the synthesized
type of the arguments $z$ to \texttt{pair}.

Meta-variables never occur in a judgment formed by $\decdir$, only in
the judgments of \reffig{dec}. In particular, these rules enforce that
meta-variables in a partial elaboration $p$ can occur \textit{only} as
type arguments in its spine, not within its head or term
arguments. This restriction guarantees \textit{spine-local}
type-argument inference and helps to narrow the programmer's focus
when debugging type errors. Furthermore, meta-variables correspond to
omitted type arguments \textit{injectively}, significantly simplifying
the kind of reasoning needed for debugging type errors. We make this
precise by defining meta-language function $MV(\Gamma,\_)$ which
yields the set of meta-variables occurring in its second argument with
respect to the context $\Gamma$. $MV$ is overloaded to take both types
and elaborated terms for its second argument: for types we define
$MV(\Gamma,T) = FV(T) - DTV(\Gamma)$, the set of free variables in $T$
less the declared type variables of $\Gamma$; for terms,
$MV(\Gamma,p)$ is defined recursively by the following equations:
\begin{alignat*}{4}
  MV(\Gamma,p)      & = \varnothing
                    & \text{ when } \neg App(p)
 \\ MV(\Gamma,p[X]) & = MV(\Gamma,p) \cup \{X\}
                    & \text{ when } X  \notin DTV(\Gamma)
 \\ MV(\Gamma,p[S]) & = MV(\Gamma,p)
                    & \text{ when } WF(\Gamma,S)
  \\ MV(\Gamma,p\ e) & = MV(\Gamma,p)
\end{alignat*}

Using our running example where the subject $t$ is $\texttt{pair}\
(\abs{\lambda}{x}{x})\ z$ we can now show how the meta-variable checks
are used in rules $AppSyn$ and $AppChk$. We have for our partially
elaborated term that $MV(\Gamma,\texttt{pair}[X][ℕ]\
(\aabs{λ}{x}{ℕ}{x})\ z) = \{X\}$ and also for our type that
${MV(\Gamma,\pair{X}{ℕ})} = \{X\}$. If we have a derivation of the
judgment above formed by $\shim$ we can then derive with rule $AppChk$
\begin{align*}
\Gamma \decchk t : ⟨(ℕ → ℕ) \times ℕ⟩ \elab \texttt{pair}[ℕ → ℕ][ℕ]\
  (\aabs{\lambda}{x}{ℕ}{x})\ z)
\end{align*}
\noindent because substitution $[ℕ \to ℕ/X]$ solves the remaining
meta-variable $X$ in the elaborated term and type, and when utilized
on the partially inferred type $⟨X \times ℕ⟩$ yields the contextual
type for the term. However, we would not be able to derive with rule
$AppSyn$
\begin{align*}
\Gamma \decsyn t : ⟨(ℕ → ℕ) \times ℕ⟩ \elab \texttt{pair}[ℕ → ℕ][ℕ]\
  (\aabs{\lambda}{x}{ℕ}{x})\ z)
\end{align*}
\noindent since we do not have $\subid$ as our solution and we have
meta-variable $X$ remaining in our partial elaboration and
type. Together, the checks in $AppSyn$ and $AppChk$ ensure that
meta-variables are never passed up and out of a maximal application
during type inference.

\paragraph{Specification Rules}
Judgment $\shim$ serves as an interface to spine-local type-argument
inference. In \reffig{decshim} it is defined in terms of the
specification for contextual type-argument inference given by
judgments $\meta$ and $\metaapp$; we call it a ``shim'' judgment
because in \reffig{algshim} we give for it an alternative
definition using the algorithmic rules in which the condition
$MV(\Gamma,T) = dom(\sigma)$ is not needed. Its purpose, then,
is to cleanly delineate what we consider specification and
implementation for our inference system.

Though the details of maintaining spine-locality and performing
synthetic type-argument inference permeate the inference rules for
$\meta$ and $\metaapp$, these rules form a specification in that they
fully abstract away the details of contextual type-argument inference,
describing how solutions are used but omitting how they are generated.
Spine-locality in particular contributes to our specification's
perceived complexity -- what would be one or two rules in a
fully-uncurried language with all-or-nothing type argument
applications is broken down in our system in to multiple inference
rules to support currying and partial type applications.

Judgment $\meta$ contains three rules and serves to dig through a
spine until it reaches its head, then work back up the spine typing
its term and type applications. The reading for it is the same as for
$\shim$, less the optional contextual type. Rule $PHead$ types the
spine head $t$ by deferring to $\decsyn$; our partial solution is
$\subid$ since no meta-variables are present in a judgment formed by 
$\decsyn$. $PTApp$ is similar to $TApp$ except it additionally
propagates solution $\sigma$. Rule $PApp$ is used for term
applications: first it partially synthesizes a type for the applicand
$t$ and then it uses judgment $\metaapp$ to ensure that the elaborated
term $p$ with this type can be applied to argument $t'$.

Judgment $\metaapp$ performs synthetic and contextual type-argument
inference and ensures that term applications with omitted type
arguments are well-typed. We read ${\Gamma \metaapp
(\ann{p}{T},\sigma) \cdot t' : T' \elab (p',\sigma')}$ as ``under
context $\Gamma$, elaborated applicand $p$ of partial type $T$
together with solution $\sigma$ can be applied to term $t'$; the
application has type $T'$ and elaborates $p'$ with solution $\sigma'$.''

Contextual type-argument inference happens in rule $PForall$, which
says that when the applicand has type $\abs{\forall}{X}{T}$ we can
choose to guess any well-formed $S$ for our contextual type argument
by picking $\sigma'' = [S/X]\circ\sigma$ (indicating $\sigma''$
contains all the mappings present in $\sigma$ and an additional
mapping $S$ for $X$), or choose to attempt to synthesize it later from
an argument by picking $\sigma'' = \sigma$. The details of which $S$
to guess, or whether we should guess at all, are not present in this
specificational rule. In both cases, we elaborate the applicand to
$p[X]$ of type $T$ and check that it can be applied to $t'$ -- we do
this even when we guess $S$ for $X$ to maintain the invariant that for
all elaborations $p$ and solutions $\sigma$ generated from the rules
in Figures \ref{fig:meta} and \ref{fig:metaapp} we have $dom(\sigma)
\subseteq MV(\Gamma,p)$, which we need when checking in the
(specificational) rule for $\shim$ that these guessed solutions are
ultimately justified by the contextual type (if any) of our maximal
application.

We illustrate the use of $PForall$ with an example: if the input
presented to judgment $\metaapp$ is
\begin{align*}
{(\ann{\texttt{pair}}{\abs{\forall}{X}{\abs{\forall}{Y}{X → Y → ⟨X
  \times Y⟩}}},\subid)\cdot(\abs{\lambda}{x}{x})}
\end{align*}
\noindent then after two uses of rule $PForall$
where we guess $ℕ → ℕ$ for $X$ and decline to guess for $Y$ we would
generate:
\begin{align*}
{(\ann{\texttt{pair}[X][Y]}{X \to Y \to ⟨X \times
    Y⟩},[ℕ → ℕ/X])\cdot(\abs{\lambda}{x}{x})}
\end{align*}

After working through omitted type arguments, $\metaapp$ requires that
we eventually reveal some arrow type $S \to T$ to type a term
application. When it does we have two cases, handled resp. by $PChk$
and $PSyn$: either the domain type $S$ of applicand $p$ together with
solution $\sigma$ provide enough information to fully know the
expected type for argument $t'$ (i.e. $MV(\Gamma,\sigma\ p) = \nada$),
or else they do not and we have some non-empty set of unsolved
meta-variables $\vars{Y}$ in $S$ corresponding to type arguments we
must synthesize. Having full knowledge, in $PChk$ we check $t'$ has
type $\sigma\ S$; otherwise, in $PSyn$ we try to solve meta-variables
$\vars{Y}$ by synthesizing a type for $t'$ and checking it is
instantiation $\vars{[U/Y]}$ (vectorized notation for the simultaneous
substitution of types $\vars{U}$ for $\vars{Y}$) of $\sigma\ S$. Once
done, we conclude with result type $\vars{[U/Y]}\ T$ and elaboration
$(\vars{[U/Y]}\ p)\ e$ for the application, as the meta-variables
$\vars{Y}$ of $p$ corresponding to omitted type arguments have now
been fully solved by type-argument synthesis. Together, $PChk$ and
$PSyn$ prevent meta-variables from being passed down to term argument
$t'$, as we require that it either check against or synthesize a
well-formed type.

We illustrate the use of rule $PSyn$ with and example: suppose that
under context $\Gamma$ the input presented to judgment $\metaapp$ is
\begin{align*} {(\ann{\texttt{pair}[X][Y]\
(\aabs{\lambda}{x}{ℕ}{x})}{Y \to ⟨X \times Y⟩},[ℕ → ℕ/X])\cdot z}
\end{align*}
\noindent and furthermore that $\Gamma \decsyn z : ℕ$. Then we have
instantiation $[ℕ/Y]$ from synthetic type-argument inference and use
it to produce for the application the result type $[ℕ/Y]\ ⟨X \times Y⟩
= ⟨X \times ℕ⟩$ and the elaboration $\texttt{pair}[X][ℕ]\
(\aabs{\lambda}{x}{ℕ}{x})\ z$. Note that synthesized type arguments
are used \textit{eagerly}, meaning that the typing information
synthesized from earlier arguments can in some cases be used to infer
the types of later arguments in \textit{checking} mode (see
\refsec{examples}). This is reminiscent of \textit{greedy}
type-argument inference for type systems with subtyping\cite{Ca97,
Du09}, which is known to cause unintuitive type inference failures due
to sub-optimal type arguments (i.e. less general wrt to the subtyping
relation) being inferred. As System F lacks subtyping, this problem
does not affect our type inference system and we can happily utilize
synthesized type arguments eagerly (see \refsec{discussion}).

\subsection{Soundness, Weak Completeness, and Annotation Requirements}
\labsec{annreq}
The inference rules in \reffig{dec} for our external language are
\textit{sound} with respect to the typing rules for our internal
language (i.e. explicitly typed System F), meaning that
elaborations of typeable external terms are typeable at the same
type\footnote{A complete list of proofs
for this paper can be found in the proof appendix at TODO}:
\begin{theorem}{(Soundness of $\ \decdir$):}
  \labthm{sound-dec}

  If $\ \Gamma \decdir t : T \elab e$ then $\ \Gamma \df e : T$.
\end{theorem}
Our inference rules also enjoy a trivial form of completeness that
serves as a sanity-check with respect to the internal language: since
any term $e$ in the internal language (i.e., any fully annotated term)
is also in the external language, we expect that $e$ should be typable
using the typing rules for external terms:
\begin{theorem}{(Trivial Completeness of $\ \decdir$): }
  \labthm{triv-complete-dec}

  If $\ \Gamma \df e : T$ then $\Gamma \decdir e : T \elab e$
\end{theorem}
A more interesting form of completeness comes from asking which
external terms can be typed -- after all, this is precisely what a
programmer needs to know when trying to debug a type inference
failure! Since our external language contains terms without any
annotations and our type language is impredicative System F, we know
from \cite{We98} that type inference is in general
undecidable. Therefore, to state a completeness theorem for type
inference we must first place some restrictions on the set of external
terms that can be the subject of typing.

We start by defining what it means for $t$ to be a \textit{partial
erasure} of internal term $e$. The grammar given in \refsec{syntax}
for the external language does not fully express where we hope our
inference rules will restore missing type information. Specifically,
the rules in Figures \ref{fig:bidir} and \ref{fig:dec} will try to infer annotations on bare
λ-abstractions and only try to infer missing type arguments that occur
in the applicand of a term application. For example, given (well-typed) internal
term $x [S_1] [S_2]\ y [T]$ and external term $x\ y$, our inference
rules will try to infer the missing type arguments $S_1$ and $S_2$ but
\textit{will not} try to infer the missing $T$.

A more artificial restriction on partial erasures is that the sequence
of type arguments occurring between two terms in an application can
only be erased in a right-to-left fashion. For example, given internal
term $x [S_1] [S_2]\ y [T_1] [T_2]\ z$, the external term $x\ y [T_1]\
z$ is a valid erasure ($S_1$ and $S_2$ are erased between $x$ and $y$,
and between $y$ and $z$ rightmost $T_2$ is erased), but term $x [S_2]\
y [T_2]\ z$ is not. This restriction helps preserve soundness of the
external type inference rules by ensuring that every explicit type
argument preserved in an erasure of an internal term $e$ instantiates
the same type variable it did in $e$; it is artificial because we
could instead have introduced notation for ``explicitly erased'' type
arguments in the external language, such as $x [\_] [S_2]\ y$, to
indicate the first type argument has been erased, but did not to
simplify the presentation of our inference rules and language.

The above restrictions for partial erasure are made precise by the
functions $\lowerdec{\_}$ and $\lowerdec{\_}_a$ which map an internal
term $e$ to sets of partial erasures $\lowerdec{e}$. They are defined
mutually recursively below:
\begin{alignat*}{4}
  \lowerdec{\aabs{\lambda}{x}{T}{e}}
  & = \{\aabs{\lambda}{x}{T}{t} \mid t \in \lowerdec{e}\}
  \cup \{\abs{\lambda}{x}{t} \mid t \in \lowerdec{e}\}
  \\ \lowerdec{\abs{\Lambda}{X}{e}}
  & = \{\abs{\Lambda}{X}{t} \mid t \in \lowerdec{e}\}
  \\ \lowerdec{e\ e'}
  & = \{t\ t' \mid t \in \lowerdec{e}_a \land t' \in \lowerdec{e'}\}
  \\ \lowerdec{e[S]}
  & = \{t [S] \mid t \in \lowerdec{e}\}
  \\
  \\ \lowerdec{e[S]}_a
  & = \{t \mid t \in \lowerdec{e}_a\} \cup
  \{t[S] \mid t \in \lowerdec{e}\}
  \\ \lowerdec{e}_a
  & = \lowerdec{e} \text{ otherwise }
\end{alignat*}
We are now ready to state a weak completeness theorem for typing terms
in the external language which over-approximates the annotations
required for type inference to succeed (we write
$\abs{\forall}{\vars{X}}{T}$ to mean some number of type
quantifications over type $T$)
\begin{theorem}{(Weak completeness of $\ \decsyn$):}
  \labthm{completeDF}

  Let $e$ be a term of the internal language and $t$ be a
  term of the internal languages such that $\ t \in \lowerdec{e}$. If
  $\ \Gamma \df e : T$ then $\Gamma \decsyn t : T \elab e$ when the
  following conditions hold for each sub-expression $e'$ of $e$,
  corresponding sub-expression $t'$ of $\ t$, and corresponding
  sub-derivation $\Gamma' \df e' : T'$ of $\ \Gamma \df e : T$:
  \begin{enumerate}
  \item If $e'\!=\!\aabs{\lambda}{x}{S}{e''}$ for some $S$ and $e''$, then
    $t'\!=\!\aabs{\lambda}{x}{S}{t''}$ for some $t'$
  \item If $e'$ occurs as a maximal term application in $e$ and if \\ $\Gamma'
    \meta t' : T'' \elab (p,\subid)$ for some $T$ and $p$, then
    $MV(\Gamma,p)\!=\!\nada$.
  \item If $e'$ is a term application and $t'\!=\!t_1\ t_2$ for some $t_1$
    and $t_2$, and if $\Gamma' \meta t_1 : T'' \elab (p,\subid)$ for some
    $T''$ and $p$, then $T''\!=\!\abs{\forall}{\vars{X}}{S_1 \to S_2}$ for
    some $S_1$ and $S_2$.
  \item If $e'$ is a type application and $t' = t''[S]$ for some $t''$ and
    $S$, and $\Gamma' \meta t'' : T'' \elab (p,\subid)$ for some $T''$
    and $p$, then $T''\!=\!\abs{\forall}{X}{S'}$ for some $S'$.
  \end{enumerate}
\end{theorem}
\refthm{completeDF} only considers synthetic type-argument inference,
and in practice condition (1) is too conservative thanks to contextual
type-argument inference. Though a little heavyweight, our weak
completeness theorem can be translated into a reasonable guide for
where type annotations are required when type synthesis
fails. Conditions (3) and (4) suggest that when the applicand of a
term or type application already partially synthesizes some type, the
programmer should give enough type arguments to at least reveal it has
the appropriate shape (resp. a type arrow or quantification). (2)
indicates that type variables that do not occur somewhere
corresponding to a term argument of an application should be
instantiated explicitly, as there is no way for synthetic
type-argument inference to do so. For example, in the expression $f\ z$ if $f$
has type $\abs{\forall}{X}{\abs{\forall}{Y}{Y \to X}}$ there is no way
to instantiate $X$ from synthesizing argument $z$. Finally, condition
(1) we suggest as the programmer's last resort: if the above advice
does not help it is because some λ-abstractions need annotations.

Note that in conditions (2), (3), and (4) we are not circularly
assuming type synthesis for sub-expressions of partial erasure $t$
succeeds in order to show that it succeeds for $t$, only that
\textit{if} a certain sub-expression can be typed \textit{then} we can make
some assumptions about the shape of its type or
elaboration. Conditions (3) and (4) in particular are a direct
consequence of a design choice we made for our algorithm to maintain
injectivity of meta-variables to omitted type arguments. As an
alternative, we could instead refine meta-variables when we know
something about the shape of their instantiation. For example, if we
encountered a term application whose applicand has a meta-variable
type $X$, we know it must have some arrow type and could refine $X$ to
$X_1 \to X_2$, where $X_1$ and $X_2$ are fresh
meta-variables. However, doing so means type errors may now require
non-trivial reasoning from users to determine why some meta-variables
were introduced in the first place.

Still, we find it somewhat inelegant that our characterization of
annotation requirements for type inference is not fully independent of
the inference system itself. For programmers using these guidelines,
this implies that there must be some way to interactively query the
type-checker for different sub-expressions of a program during
debugging. Fortunately, many programming languages offer just such a
feature in the form of a REPL, meaning that in practice this is not
too onerous a requirement to make.

\refthm{completeDF} only states when an external term will synthesize
its type, but what about when a term can be \textit{checked} against a
type? It is clear from the typing rules in \reffig{bidir} that some
terms that fail to synthesize a type may still be successfully checked
against a type. Besides typing bare λ-abstractions (which can only have their
type checked), checking mode can also reduce the annotation burden
implied by condition (2) of \refthm{completeDF}: consider again the
example $f\ z$ where $f$ has type $\abs{\forall}{X}{\abs{\forall}{Y}{Y
    \to X}}$. If instead of attempting type synthesis we were to check that
it has some type $T$ then we would not need to provide an explicit
type argument to instantiate $X$.

From these observations and our next result, we have that checking
mode of our type inference system can infer the types of strictly more
terms than can synthesizing mode -- whenever a term synthesizes a
type, it can be checked against the same type.
\begin{theorem}{(Checking extends synthesizing): }
  \labthm{cchksyn}

  If $\Gamma \decsyn t : T \elab e$ then $\Gamma \decchk t : T \elab e$
\end{theorem}
\subsection{Examples}
\labsec{examples}
\begin{figure*}
  \begin{subfigure}{1\linewidth}
    \centering
    \[
    \infer[AppChk]
    {
      \Gamma \decchk \texttt{pair}\ (\abs{\lambda}{x}{x})\ z : ⟨(ℕ → ℕ) \times
      ℕ⟩ \elab \texttt{pair}[ℕ \to ℕ][ℕ]\ (\aabs{\lambda}{x}{ℕ}{x})\ z
    }
    {
      \infer
      {
        \Gamma \shim \texttt{pair}\ (\abs{\lambda}{x}{x})\ z : ⟨X \times
        ℕ⟩ \elab (p,\sigma)
      }
      {\infer[PApp]
      { \Gamma \meta \texttt{pair}\ (\abs{\lambda}{x}{x})\ z : ⟨X \times
        ℕ⟩ \elab (p,\sigma)
      }
      {
        \infer[PApp]
        { \Gamma \meta \texttt{pair}\ (\abs{\lambda}{x}{x})\ : Y \to ⟨X \times
          ℕ⟩ \elab (\texttt{pair}[X][ℕ]\ (\aabs{\lambda}{x}{ℕ}{x}),\sigma)}
        {
          \infer[PHead]
          {
            \Gamma \meta \texttt{pair} :
            \abs{\forall}{X}{\abs{\forall}{Y}{I_{\times}\ X\ Y}} \elab (\texttt{pair},\subid)
          }
          {
            \infer[Var]
            { \Gamma \decsyn \texttt{pair} :
              \abs{\forall}{X}{\abs{\forall}{Y}{I_{\times}}\ X\ Y}
              \elab \texttt{pair} }
            {\ }
        }
          & \mathcal{D}_1
        }
        & \mathcal{D}_2
      }
      & MV(\Gamma,X \times ℕ) = dom(\sigma)
    }
      &  MV(\Gamma,p) = dom(\sigma)
    }
    \]
  \end{subfigure}
  \begin{subfigure}{1\linewidth}
    \ \\ \ \\ 
    \[ \mathcal{D}_1 = 
      \infer[PForall]
      { \Gamma \metaapp
        (\ann{\texttt{pair}}{\abs{\forall}{X}{\abs{\forall}{Y}{I_{\times}\
              X\ Y}}},\subid) \cdot (\abs{\lambda}{x}{x}) : Y \to ⟨X
        \times Y⟩ \elab (\texttt{pair}[X][Y]\ (\lamnatid),\sigma)}
      {
        \infer[PForall]
        { \Gamma \metaapp
          (\ann{\texttt{pair}[X]}{\abs{\forall}{Y}{I_{\times}\ X\
              Y}},\sigma) \cdot (\lamid) : Y \to ⟨X
          \times Y⟩ \elab (\texttt{pair}[X][Y]\ (\lamnatid),\sigma)
        }
        {
          \infer[PChk]
          {
            \Gamma \metaapp (\ann{\texttt{pair}[X][Y]}{I_{\times}\ X\ Y},\sigma) \cdot (\lamid) : Y \to ⟨X
            \times Y⟩ \elab (\texttt{pair}[X][Y]\ (\lamnatid),\sigma)
          }
          {
            MV(\Gamma,\sigma\ X) = \nada
            && \infer[Abs]
            { \Gamma \decchk \lamid : ℕ \to ℕ \elab \lamnatid}
            { \infer[Var]
              { \Gamma,\ann{x}{ℕ} \decchk x : ℕ \elab x}
              {\ }
            }
          }
        }
      }
    \]
  \end{subfigure}
  \begin{subfigure}{1\linewidth}
    \ \\ \ \\
    \[ \mathcal{D}_2 = \infer[PSyn]
      {
        \Gamma \metaapp (\ann{\texttt{pair}[X][Y]\ (\lamnatid)}{Y \to
          ⟨X \times Y⟩},\sigma)
        \cdot z : ⟨X \times ℕ⟩ \elab (\texttt{pair}[X][ℕ]\
        (\lamnatid)\ z,\sigma)
      }
      {
        MV(\Gamma,Y) = \{Y\}
        & \infer[Var]
        { \Gamma \decsyn z : ℕ \elab z}
        { \ }
      }
    \]
  \end{subfigure}
  \ \\ \ \\
  \begin{eqnarray*}
    \text{where:\ \ \ \ }
    I_{\times}\ X\ Y & = & \introx{X}{Y}
    \\\ \Gamma & = &
                    \ann{\texttt{pair}}{\abs{\forall}{X}{\abs{\forall}{Y}{I_{\times}\
                    X\ Y}}},\ \ann{z}{ℕ}
    \\ \sigma & = & [ℕ → ℕ/X]
    \\ p      & = & \texttt{pair}[X][ℕ]\ (\lamnatid)\ z
  \end{eqnarray*}
  \caption{Example typing derivation with the specification rules}
  \labfig{ex1}
\end{figure*}
\paragraph{\textbf{Successful Type Inference}}
We conclude this section with some example programs for which the type
inference system in Figures \ref{fig:bidir} and \ref{fig:dec} will and
will not be able to type. We start with the motivating example from
the introduction of checking that the expression $\texttt{pair}\
(\lamid)\ z$ has type $⟨(ℕ → ℕ) \times ℕ⟩$, which is not possible in
other variants of local type inference. For convenience, we assume the
existence of a base type ℕ and a family of base types $\pair{S}{T}$
for all types $S$ and $T$. These assumptions are admissible as we
could define these types using Church encodings. A full derivation for
typing this program is given in \reffig{ex1}, including the following
abbreviations:
\begin{eqnarray*}
  I_{\times}\ X\ Y & = & \introx{X}{Y}
  \\\ \Gamma & = &
                   \ann{\texttt{pair}}{\abs{\forall}{X}{\abs{\forall}{Y}{I_{\times}\
                   X\ Y}}},\ \ann{z}{ℕ}
  \\ \sigma & = & [ℕ → ℕ/X]
  \\ p      & = & \texttt{pair}[X][ℕ]\ (\lamnatid)\ z
\end{eqnarray*}
To type this application ${\texttt{pair}\ (\lamid)\ z}$ we first dig
through the spine, reach the head \texttt{pair}, and synthesize type
$\abs{\forall}{X}{\abs{\forall}{Y}{\texttt{I}_\times\ X\ Y}}$. No
meta-variables are generated by judgment $\decsyn$ and thus there can
be no meta-variable solutions, so we generate solution $\subid$.

Next we type the first application, $\texttt{pair}\ (\lamid)$, shown
in sub-derivation $\mathcal{D}_1$. In the first invocation of rule
$PForall$ we guess solution $\sigma$ for $X$, and in the second
invocation we decline to guess an instantiation for $Y$ (in this
example we could have also guessed $\mathbb{N}$ for $Y$ as this
information is also available from the contextual type, but choose not
to in order to demonstrate the use of all three rules of
$\metaapp$). Then using rule $PChk$ we check argument
$\lamid$ against $\sigma\ X = \mathbb{N} \to \mathbb{N}$. \textit{This
is the point at which the local type inference systems of
\cite{PT98,OZZ01} will fail}: as a bare λ-abstraction this argument
will not synthesize a type, and the expected type $X$ as provided by the
applicand \texttt{pair} alone does not tell us what the missing type
annotation should be. However, by using the information provided by the
contextual type of the entire application we know it must
have type $ℕ \to ℕ$.  The resulting partial type of the application is
$Y \to \pair{X}{Y}$, and we propagate solution $\sigma$ to the rest of
the derivation. Note that we elaborate the argument $\lamid$ of this
application to $\lamnatid$ -- we never pass down meta-variables to
term arguments, keeping type-argument inference local to the spine.

In sub-derivation $\mathcal{D}_2$ we type $(\texttt{pair}\
(\abs{\lambda}{x}{x}))\ z$ (parentheses added) where our applicand has
partial type $Y \to \pair{X}{Y}$. We find that we have unsolved
meta-variable $Y$ as the expected type for $z$, so we use rule $PSyn$
and synthesize the type $\mathbb{N}$ for $z$. Using solution $[ℕ/Y]$,
we produce $\pair{X}{Y}$ for the resulting type of the application and
elaborate the application to a $\texttt{pair}[X][ℕ]\ (\lamnatid)\ z$,
wherein type argument $Y$ is replaced by ℕ in the original elaborated
applicand $\texttt{pair}[X][Y]\ (\lamnatid)$.

Finally, in rule $AppChk$ we confirm that the only meta-variables
remaining in our partial type synthesis of the application is
precisely those for which we knew the solutions from the contextual
type. For this example, the only remaining meta-variable in both the
partially synthesized type and elaboration is $X$, which is also the
only mapping in $\sigma$, so type inference succeeds. We use $\sigma$
to replace all occurrences of $X$ with $ℕ \to ℕ$ in the type and
elaboration and conclude that term $\texttt{pair}\ (\lamid)\ z$ can be
checked against type $\pair{(ℕ → ℕ)}{ℕ}$.

The next example illustrates how our eager use of \textit{synthetic}
type-argument inference can type some terms not possible in other
variants of local type inference. Consider checking that the
expression $\texttt{rapp}\ x\ \abs{\lambda}{y}{y}$ has type ℕ, where
$\texttt{rapp}$ has type $\abs{\forall}{X}{\abs{\forall}{Y}{X \to (X
\to Y) \to Y}}$ and $x$ has type ℕ. From the contextual type we know
that $Y$ should be instantiated to ℕ, and when we reach application
$\texttt{rapp}\ y$, we learn that $X$ should be instantiated to ℕ from
the synthesized type of $y$. Together, this gives us enough
information to know that argument $\abs{\lambda}{y}{y}$ should have type $ℕ \to
ℕ$. Such eager instantiation is neither novel nor necessarily
desirable when extended to richer types or more powerful systems of
inference (see \refsec{discussion}), but in our setting it is a useful
optimization that we happily make for inferring the types of expressions
like the one above.

\paragraph{\textbf{Type Inference Failures}}
To see where type inference can fail, we again use $\texttt{pair}\
(\abs{\lambda}{x}{x})\ z$ but now ask that it \textit{synthesize} its
type. Rule $AppSyn$ insists that we make no guesses for
meta-variables (as there is no contextual type for
the application that they could have come from), so we would need to
synthesize a type for argument $\abs{\lambda}{x}{x}$ -- but our rules
do not permit this! In this case the user can expect an error message
like the following:
\begin{verbatim}
expected type: ?X
        error: We are not in checking mode, so bound 
               variable x must be annotated
\end{verbatim}
\noindent where \verb;?X; indicates an unsolved meta-variable
corresponding to type variable $X$ in the type of \texttt{pair}. The
situation above corresponds to condition (1) of \refthm{completeDF}:
in general, if there is not enough information from the type of an
applicand and the contextual type of the application spine in which it
occurs to fully know the expected types of arguments that are
λ-abstractions, then such arguments require explicit type annotations.

We next look at an example corresponding to condition (2) of
\refthm{completeDF}, namely that the type variables of a polymorphic
function that do not correspond to term arguments in an application
should be instantiated explicitly. Here we will assume a
family of base types $S + T$ for every type $S$ and $T$, a variable
\texttt{right} of type $\abs{\forall}{X}{\abs{\forall}{Y}{Y \to (X +
    Y)}}$, and a variable $z$ of type $ℕ$. In trying to synthesize a
type for the application $\texttt{right}\ z$ the user can expect
an error message like:
\noindent
\begin{verbatim}
synthesized type: (?X + ℕ)
              error: This maximal application has unsolved
                  meta-variables
\end{verbatim}
\noindent indicating that type variable $X$ requires an explicit type
argument be provided. Fortunately for the programmer, and unlike the
local type inference systems of \cite{PT98,OZZ01}, our system supports
partial explicit type application, meaning that $X$ can be
instantiated without also explicitly (and redundantly) instantiating
$Y$. On the other hand, local type inference systems for System
F$_\leq$\cite{PT98,OZZ01} can succeed to type $\texttt{right}\ z$
\textit{without} additional type arguments, as they can instantiate
$X$ to the minimal type (with respect to their subtyping relation)
\texttt{Bot}. Partial type application, then, is more useful for our
setting of System F where picking some instantiation for this
situation would be somewhat arbitrary.

A more subtle point of failure for our algorithm corresponds to
conditions (3) and (4) of \refthm{completeDF}. Even when the head
and all arguments of an application spine can synthesize their types,
the programmer may still be require to provide some additional type
arguments. Consider the expression ${\texttt{bot}\ \texttt{z}}$, where
$\ann{\texttt{bot}}{\abs{\forall}{X}{X}}$ and
$\ann{\texttt{z}}{\mathbb{N}}$. Even with some contextual type for this
expression, type inference still fails because the
rules in \reffig{metaapp} require that the type of the applicand of a
term application reveals some arrow, which $\abs{\forall}{X}{X}$ does
not. The programmer would be met with the following error message:
\begin{verbatim}
applicand type: ?X
         error: The type of an applicand in a term
                application must reveal an arrow
\end{verbatim}
\noindent prompting the user to provide an explicit type argument for
$X$. To make expression $\texttt{bot}\ z$ typeable, the programmer
could write $\texttt{bot}[ℕ → ℕ]\ z$, or even
$\texttt{bot}[\abs{\forall}{Y}{Y \to Y}]\ z$ -- our inference rules
are able to solve meta-variables introduced by explicit and even
synthetic type arguments, as long as there is at least enough
information to reveal a quantifier or arrow in the type of a term or
type applicand.

For our last type error example, we consider the situation where the
programmer has written an ill-typed program. Local type inference
enjoys the property that type errors can be understood
\textit{locally}, without any ``spooky action'' from a distant part of
the program. In particular, with local type inference we would like to
avoid error messages like the following:
\begin{verbatim}
synthesized type: 𝔹 → 𝔹
   expected type: ?X := ℕ → ℕ
           error: type mismatch
\end{verbatim}
\noindent From this error message alone the programmer has no
indication of why the expected type is $ℕ → ℕ$! In our type inference
system we expand the distance information travels by allowing
it to flow from the contextual type of an application to its
arguments. As an example, the error message above might be generated
when checking that the expression $\texttt{pair}\
(\aabs{\lambda}{x}{𝔹}{x})\ z$ has type $⟨(ℕ → ℕ) \times ℕ⟩$,
specifically when inferring the type of the first
argument. Fortunately, our notion of locality is still quite
small and we can easily demystify the reason type inference expected a
different type:
\begin{verbatim}
synthesized type: 𝔹 → 𝔹
   expected type: ?X := ℕ → ℕ
contextual match: ⟨?X × ?Y⟩ := ⟨(ℕ → ℕ) × ℕ⟩
\end{verbatim}
\noindent where \texttt{contextual match} tells the programmer to
compare to the partially synthesized and contextual return types of the
application to determine why $X$ was instantiated to $ℕ \to ℕ$. A
similar field, \texttt{synthetic match}, could tell the programmer
that the type of an earlier argument informs the expected type of current one.
\section{Algorithmic Inference Rules}
\labsec{arules}
\begin{figure*}
  \begin{subfigure}{1\linewidth}
    \centering
    \caption{Shim (algorithm)}
    \labfig{algshim}
    \[
      \begin{array}{cc}
      T_? ::= T\ |\ ?
      & \infer
      {
        \Gamma ; T_? \shim t\ t' : T \elab (p,\sigma)
      }
      {
        \Gamma ; T_? \proto t\ t' : T \elab (p,\sigma)
        }
      \end{array}
    \]
  \end{subfigure}
  \begin{subfigure}{1\linewidth}
    \centering
    \caption{\fbox{$\Gamma;P \proto t : W \elab (p,\sigma)$}}
  \[
    \begin{array}{cc}
      \infer[?Head]
      {\Gamma ; ? \to P \proto t : W \elab (e, \subid)}
      { \neg App(t)
        & \Gamma \algsyn t : T \elab e
        & \varnothing \match T :=\ ? \to P \To(\subid,
          W) }
      & \infer[?TApp]
      { \Gamma ; ? \to P \proto t[S] : [S/X]W \elab (p[S], \sigma)}
      { \Gamma ; ? \to P \proto t : \abs{\forall}{\dec{X}{R}}{W} \elab (p,\sigma)
      & R \in \{X,S\} }
      \\ \\
      \infer[?App]
      {\Gamma ; P \proto t\ t' : W' \elab (p', \sigma') }
      { \Gamma ; ? \to P \proto t : W \elab (p, \sigma)
        & \Gamma \algapp (\ann{p}{W},\sigma) \cdot t' : W' \elab
          (p',\sigma')}
   \end{array}
 \]
 \labfig{proto}
\end{subfigure}
\begin{subfigure}{1\linewidth}
  \centering
  \caption{\fbox{$\Gamma \algapp (\ann{p}{W},\sigma)\cdot t' : W \elab
    (p',\sigma')$}}
  \[
    \begin{array}{cc}
      \infer[?Forall]
      { \Gamma \algapp (\ann{p}{\abs{\forall}{\dec{X}{R}}}{W},\sigma) \cdot t' : W' \elab (p', \sigma')}
      { \sigma'' = \text{ if } R\!=\!X \text{ then } \sigma \text { else }
        [R/X]\!\circ\!\sigma
        & \Gamma \algapp (\ann{p[X]}{W},\sigma'') \cdot t' : W' \elab (p', \sigma')
      }
      & \infer[?Chk]
        {\Gamma \algapp (\ann{p}{S \to W},\sigma) \cdot t' : W \elab (p\ e', \sigma)}
        { MV(\Gamma, \sigma\ S) = \varnothing
          & \Gamma \algchk t' : S \elab e }
      \\ \\
      \infer[?Syn]
      { \Gamma \algapp (\ann{p}{S \to W}) \cdot t' : \vars{[U/Y]}\ W
      \elab ((\vars{[U/Y]}\ p)\ e', \sigma)}
      { MV(\Gamma, \sigma\ S) = \vars{Y} \neq \varnothing
        & \Gamma \algsyn t : \vars{[U/Y]}\ \sigma\ S \elab e
      }
    \end{array}
  \]
  \labfig{algapp}
\end{subfigure}
\begin{subfigure}{1\linewidth}
  \centering
  \caption{\fbox{$\vars{X} \match T := P \To (\sigma,W)$}}
  \[
    \begin{array}{ccc}
      \infer[MArr]
      {\vars{X} \match S \to T :=\ ?\!\to P \To (\sigma, S \to W)}
      {\vars{X} \match T := P \To (\sigma, W)}
      & \infer[MType]
        {\vars{X} \match T := S \To (\vars{[U/X]}, T)}
        { \vars{[U/X]}\ T = S }
      & \infer[M?]
        {\vars{X} \match T :=\ ? \To (\subid, T)}
        {}
    \end{array}
    \] \\ \[
      \begin{array}{cc}
      \infer[MForall]
        {\vars{X} \match \abs{\forall}{X}{T} :=\ ?\!\to P \To
        (\sigma - X, \abs{\forall}{\dec{X}{\sigma(X)}}{W})}
      {\vars{X},X \match T :=\ ?\!\to P \To (\sigma, W)}
      & \infer[MCurr]
        {\vars{X} \match X :=\ ?\!\to P \To (\subid, (X, ?\!\to P))}
        {X \in \vars{X}}
    \end{array}
  \]
  \labfig{pmatch}
\end{subfigure}
\caption{Algorithm for contextual type argument inference}
\labfig{alg}
\end{figure*}
The type inference system presented in \refsec{drules} do not
constitute an algorithm. Though the rules forming
judgment $\metaapp$ indicate \textit{where} and \textit{how} we use
contextually-inferred type arguments, they do not specify 
\textit{what} their instantiations are or even \textit{whether} this
information is available to use, and it is not obvious how to work
backwards from the second premise in \reffig{decshim} to develop an
algorithm.

\reffig{alg} shows the algorithmic rules implementing contextual
type-argument inference. The full algorithm for spine-local type
inference, then, consists of the rules in \reffig{bidir} with the shim
judgment $\shim$ as defined in \reffig{algshim}. At the heart of our
implementation is our prototype matching algorithm; to understand the
details of how we implement contextual type-argument inference, we must
first discuss this algorithm and the two new syntactic categories it
introduces, prototypes and decorated types.

\subsection{Prototype Matching}

\reffig{pmatch} lists the rules for the prototype matching
algorithm. We read the judgment $\vars{X} \match T := P \To
(\sigma,W)$ as: ``solving for meta-variables $\vars{X}$, we match type
$T$ to prototype $P$ and generate solution $\sigma$ and decorated
type $W$,'' and we maintain the invariant that $dom(\sigma) \subseteq
\vars{X}$. Meta-variables can only occur in $T$, thus these are
\textit{matching} (not unification) rules. The grammar for
prototypes and decorated types is given below:
\begin{align*}
\textbf{Prototypes}      && P & ::=\ ?\ |\ T\ |\ \parr P \\
\textbf{Decorated Types} && W & ::= T\ |\ S \to W\ |\
                                \dabs{X}{X}{W}\ |\
                                \dabs{X}{S}{W} \\
                         &&   & \ \ \ \ \ |\ (X,\parr P)
\end{align*}

Prototypes carry the contextual type of the maximal application of a
spine. In the base case they are either the uninformative $?$ (as in
$AppSyn$), indicating no contextual type, or they are informative of
type $T$ (as in $AppChk$). In this way, prototypes generalize the
syntactic category $T_?$ we introduced earlier for optional contextual
types. We use the last prototype former $\parr$ as we work our way
down an application spine to track the expected arity of its head. For
example, if we wished to check that the expression \verb;id suc x; has
type $\mathbb{N}$, then when we reached the head \verb;id; using the
rules in \reffig{proto} we would generate for it prototype $\parr\
\parr \mathbb{N}$

Decorated types consist of types (also called \textit{plain-decorated}
types), an arrow with a regular type as the domain (as prototypes only
inform us of the result type of a maximal application, not of the
types of arguments), quantified types whose bound variable $X$ may be
decorated with the type to which we expect to instantiate it, and ``stuck''
decorations. On quantifiers, decoration $\dec{X}{X}$ indicates that
$P$ did not inform us of an instantiation for $X$ -- we sometimes
abbreviate the two cases as $\dabs{X}{R}{W}$, where $R \in \{X,S\}$
and $S \neq X$.

To explain the role of stuck decorations, consider again \verb;id suc x;.
Assuming \verb;id; has type $\abs{\forall}{X}{X \to X}$, matching
this with prototype $\parr\ \parr \mathbb{N}$ generates decorated type
$\dabs{X}{X}{X \to (X,\parr \mathbb{N})}$, meaning that we only know
that $X$ will be instantiated to some type that matches $\parr
\mathbb{N}$. Stuck decorations occur when the expected arity of a
spine head (as tracked by a given prototype) is greater than the arity
of the type of the head and are the mechanism by which we propagate a
contextual type to a head that is ``over-applied'' -- a not-uncommon
occurrence in languages with curried applications!

Turning to the prototype matching algorithm in \reffig{pmatch}, rule
$MArr$ says that we match an arrow type and prototype when we can
match their codomains. Rule $MType$ says that when the prototype is
some type $S$ we must find an instantiation $\vars{[U/X]}$ such that
$\vars{[U/X]}\ T = S$, and rule $M?$ says that any type matches with
$?$ with no solutions generated (thus we call $?$ the
``uninformative'' prototype). In rule $MForall$ we match a quantified
type with a prototype by adding bound variable $X$ to our
meta-variables and matching the body $T$ to the same prototype; the
substitution in the conclusion, $\sigma - X$, is the solution
generated from this match less its mapping for $X$, which is placed in
the decoration $\dec{X}{\sigma(X)}$. For example, matching
$\abs{\forall}{X}{\abs}{\forall}{Y}{X \to Y \to X}$ with prototype
$\parr \parr ℕ$ generates decorated type ${\dabs{X}{ℕ}{\dabs{Y}{Y}{X
\to Y \to X}}}$. Finally, rule $MCurr$ applies when there is incomplete
information (in the form of $\ \parr\ P$) on how to instantiate a
meta-variable; we generate a stuck decoration with identity solution
$\subid$.

We conclude by showing that our prototype matching rules really do
constitute an algorithm: when $\vars{X}$, $T$, and $P$ are considered
as inputs then $\match$ behaves like a function.

\begin{theorem}{(Function-ness of $\ \match$): }

  Given $\vars{X}$, $T$, and $P$, if $\ \vars{X} \match T := P \To
  (\sigma,W)$

  and ${\vars{X} \match T := P \To (\sigma',W')}$, then
  $\sigma = \sigma'$ and $W = W'$
\end{theorem}

\subsection{Decorated Type Inference}

We now discuss the rules in Figures \ref{fig:proto} and
\ref{fig:algapp} which implement contextual type-argument inference
(as specified by Figures \ref{fig:meta} and \ref{fig:metaapp}) by
using the prototype matching algorithm. We begin by giving a reading
for judgments $\proto$ -- read ${\Gamma;P \proto t : W \elab
(p,\sigma)}$ as: ``under context $\Gamma$ and with prototype $P$, $t$
synthesizes decorated type $W$ and elaborates $p$ with solution
$\sigma$,'' where $\sigma$ again represents the contextually-inferred
type arguments.

In rule $AppSyn$ we required that the solution generated by $\shim$ in
its premise is $\subid$; in $AppChk$ we (implicitly) required that the
contextual type is equal to $\sigma\ T$; and now with the algorithmic
definition for $\shim$ we appear to be requiring in both that the decorated
type generated by $\proto$ is a plain-decorated type $T$. With the
algorithmic rules, these are not requirements but \textit{guarantees}
that the specification makes of the algorithm:
\begin{lemma}
  \labthm{arrprotodec}
  Let $arr_P(P)$ be the number of prototype arrows prefixing $P$ and
  $arr_W(W)$ be the number of decorated arrows prefixing $W$.
  If $\ {\Gamma;P \proto t : W \elab (p,\sigma)}$ then $arr_W(W) \leq arr_P(P)$
\end{lemma}
\begin{theorem}{(Soundness of $\ \proto$ wrt $\ \match$): }
  \labthm{sprotomatch}

  If $\ \Gamma;P \proto t : W \elab (p,\sigma)\ $ then $MV(\Gamma,p)
  \match T := P \To (\sigma,W)$
\end{theorem}
Assuming prototype inference succeeds, when we specialize $P$ in
\refthm{sprotomatch} to $?$ we have immediately by rule $M?$ that
$\sigma = \subid$; when we specialize it to some contextual type $T'$
for an application, then by the premise of $MType$ we have $\sigma\ T
= T'$. \refthm{arrprotodec} and \ref{thm:sprotomatch} together 
tell us that we generate plain-decorated types in both cases, as in
particular we cannot have leading (decorated) arrows or stuck
decorations with prototypes $?$ or $T'$.

Next we discuss the rules forming judgment $\proto$ in \reffig{proto},
constituting the algorithmic version of the rules in \reffig{meta}. In
rule $?Head$, after synthesizing a type $T$ for the application head
we match this type against expected prototype $\parr P$ (we are
guaranteed the prototype has this shape since only a term application
can begin a derivation of $\proto$). No meta-variables occur in $T$
initially -- as we perform prototype matching these will be generated
by rule $MForall$ from quantified type variables in $T$ and their
solutions will be left as decorations in the resulting decorated type
$W$. We are justified in requiring that matching $T$ to $\parr P$
generates empty solution $\subid$ since we have in general that the
meta-variables solved by our prototype matching judgment are a subset
of the meta-variables it was asked to solve:

\begin{lemma}
  If $\vars{X} \match T := P \To (\sigma,W)$ then $dom(\sigma)
  \subseteq \vars{X}$
\end{lemma}

In $?TApp$, we can infer the type of a type application $t[S]$ when
$t$ synthesizes a decorated type $\dabs{X}{R}{W}$ and $R$ is either an
uninformative decoration $X$ or is precisely $S$ (that is, the
programmer provided explicitly the type argument the algorithm
contextually inferred). We synthesize $[S/X]W$ for the type
application, where we extend type substitution to decorated types by the
following recursive partial function:
\begin{alignat*}{4}
  \\ \sigma\ S \to W        & = (\sigma\ S) \to (\sigma\ W)
  \\ \sigma\ \dabs{X}{R}{W} & = \dabs{X}{R}{\sigma\ W}
  \\ \sigma\ (X,\parr P)    & = W
                            & \text{if } \nada \match \sigma(X) :=
                              \parr P \To (\subid,W)
\end{alignat*}
This definition is straightforward except for the last case dealing
with stuck decorations. Here, $\sigma$ (representing instantiations
given by explicit or synthetically-inferred type arguments) may
provide information on how to instantiate $X$ and this must match our
current (though incomplete) information from $\parr P$ about our
contextually-inferred type arguments. For example,
if we have decorated type $W = X \to (X,\parr \mathbb{N})$, then
$[\mathbb{N} \to \mathbb{N}/X]\ W$ would require we match $\mathbb{N}
\to \mathbb{N}$ with $\parr \mathbb{N}$ and matching would generate
(plain) decorated type $(\mathbb{N} \to \mathbb{N}) \to \mathbb{N} \to
\mathbb{N}$

The definition of substitution on decorated types is partial
since prototype matching may fail (consider if we used substitution $[ℕ/X]$
in the above example instead). When a decorated type substitution
$\sigma\ W$ appears in the conclusion of our algorithmic rules, such as
in $?TApp$ or $?Syn$, we are implicitly assuming an additional
premise that the result is defined.

The last rule for judgment $\proto$ is $?App$, and like $PApp$ it
benefits from a reading for judgment $\algapp$ occurring in its
premise. We read $\Gamma \algapp (\ann{p}{W},\sigma) \cdot t' : W'
\elab (p',W')$ as: ``under $\Gamma$, elaborated applicand $p$ of
decorated type $W$ together with solution $\sigma$ can be
applied to $t'$; the application has decorated type $W'$ and
elaborates $p'$ with solution $\sigma'$.'' Thus, $?App$ says that to
synthesize a decorated type for a term application $t\ t'$ we
synthesize the decorated type of the applicand $t$ and ensure that the
resulting elaboration $p$, along with its decorated type and
solution, can be applied to $t'$.

We now turn to the rules for the last judgment $\algapp$ of our
algorithm. Rule $?Forall$ clarifies the non-deterministic guessing
done by the corresponding specificational rule $PForall$: the
contextually-inferred type arguments we build during contextual
type-argument inference are just the accumulation of quantified type
decorations. The solution $\sigma''$ we provide to the second premise
of $?Forall$ contains mapping $[R/X]$ if $R$ is an informative
decoration, and as we did in rule $PForall$ we provide elaborated term
$p[X]$ to track the contextually-inferred type arguments separately
from those synthetically inferred.

Rule $?Chk$ works similarly to $PChk$: when the only meta-variables in
the domain $S$ of our decorated type are solved by $\sigma$, we can
check that argument $t'$ has type $\sigma\ S$. In rule $?Syn$ we have
some meta-variables $\vars{Y}$ in $S$ not solved by $\sigma$ -- we
synthesize a type for the argument, ensure that it is some
instantiation $\vars{[U/Y]}$ of $\sigma\ S$, and use this
instantiation on the meta-variables in $p$ as well as the decorated
codomain type $W$, potentially unlocking some stuck decoration to
reveal more arrows or decorated type quantifications.

We conclude this section by noting that the specificational and
algorithmic type inference system are equivalent, in the sense that
they type precisely the same set of terms:
\begin{theorem}{(Soundness of $\algdir$ wrt $\decdir$): }
  \labthm{salgdir}

  If $\ \Gamma \algdir t : T \elab e$ then $\ \Gamma \decdir t : T
  \elab e$
\end{theorem}
\begin{theorem}{(Completeness of $\algdir$ wrt $\decdir$): }
  \labthm{calgdir}

  If $\ \Gamma \decdir t : T \elab e$ then $\ \Gamma \algdir t : T \elab
  e$
\end{theorem}
\noindent (where $\algdir$ indicates $\shim$ is defined as in
\reffig{algshim})

Taken together, Theorems \ref{thm:salgdir} and \ref{thm:calgdir}
justify our claim that the rules of \reffig{dec} constitute a
specification for contextual type-argument inference -- it is not
necessary that the programmer know the notably more complex details of
prototype matching or type decoration to understand how contextual
type arguments are inferred. Indeed, the judgment $\metaapp$ provides
more flexibility in reasoning about type inference than does
$\algapp$, as in rule $PForall$ we may freely decline to
guess a contextual type argument even when this would be justified
and instead try to learn it synthetically. In contrast, algorithmic
rule $?Forall$ requires that we use any informative quantifier
decoration. We use this flexibility when giving guidelines for the
required annotations in \refsec{annreq} for typing external terms, as
the required conditions for typeability in \refthm{completeDF} would
be further complicated if we could not restrict ourselves to using
only synthetic type-argument inference.

%% file: slti-discussion.tex
\section{Discussion \& Related Work}
\labsec{discussion}

\subsection{Local Type Inference and System F$_{\leq}$}
\paragraph{\textbf{Local Type Inference}}
Our work is most influenced by the seminal paper by Pierce and
Turner\cite{PT98} on local type inference that describes its broad
approach, including the two techniques of bidirectional typing rules
and local type-argument inference and the design-space restriction
that polymorphic function applications be fully-uncurried to maximize
the benefit of these techniques. In their system, either all term
arguments to polymorphic functions must be synthesized or else all
type arguments must be given -- no compromise is available when only a
\textit{few} type arguments suffice to type an application, be they
provided explicitly or inferred contextually. Our primary motivation
in this work was addressing these issues -- restoring first-class
currying, enabling partial type application, and utilizing the
contextual type of an application for type-argument inference -- while
maintaining some of the desirable properties of local type inference
and staying in the spirit of their approach.

\paragraph{\textbf{Colored Local Type Inference}}
Odersky, Zenger, and Zenger\cite{OZZ01} improve upon the type system
of Pierce and Turner by extending it to allow \textit{partial} type
information to be propagated downwards when inferring types for term
arguments. Their insight was to internalize the two modes of
bidirectional type inference to the structure of types themselves,
allowing different parts of a type to be synthetic or contextual. In
contrast, we use an ``all or nothing'' approach to type propagation:
when we encountered a term argument for which we have incomplete
information, we require that it fully synthesize its type. On the
other hand, their system uses only the typing information provided by
the application head, whereas we combine this with the contextual type
of an application, allowing us to type some expressions their
system cannot. The upshot of the difference in these systems is that
spine-local type inference utilizes \textit{more} contextual
information and colored local type inference utilizes contextual
information \textit{more cleverly}.

The syntax for prototypes in our algorithm was directly inspired by
the prototypes used in the algorithmic inference rules for
\cite{OZZ01}. Our use of prototypes complements theirs; ours
propagates the partial type information provided by contextual type of
an application spine to its head, whereas theirs propagates the
partial type information provided by an application head to its
arguments. In future work, we hope to combine these two notions of
prototype to propagate \textit{partially} the type information coming
from the application's contextual type \textit{and} head to its
arguments.

\paragraph{\textbf{Subtyping}}
Local type inference is usually studied in the setting of System
F$_{\leq}$ which combines impredicative parametric polymorphism and
subtyping. The reason for this is two-fold: first, a partial type
inference technique is needed as complete type inference for
F$_{\leq}$ is undecidable\cite{TU96}; second, global type inference
systems fail to infer principal types in F$_{\leq}$
\cite{Od02,OSW99,Ke96}, whereas local type inference is able to
promise that it infers the ``locally best''\cite{PT98} type arguments
(i.e. the type arguments minimizing the result type of the
application, relative to the subtyping relation). The setting for our
algorithm is System F, so the reader may ask whether our developments
can be extended gracefully to handle subtyping. We believe the answer
is yes, though with some modification on how synthetic type
arguments are used.

In rule $PSyn$ in \reffig{metaapp}, meta-variables $\vars{Y}$ are
instantiated to types $\vars{U}$ immediately. In the presence of
subtyping this would make our rules \textit{greedy}\cite{Ca97,Du09}
and we would not be able to guarantee synthetic type-argument
inference produced locally best types, possibly causing type inference
to fail later in the application spine. To illustrate this, consider
the expression $\texttt{rapp}\ x\ \texttt{neg}$, assuming $\texttt{rapp}
: \abs{\forall}{X}{\abs{\forall}{Y}{X \to (X \to Y) \to Y}}$, $x : ℕ$,
$\texttt{neg} : \mathbb{Z} \to \mathbb{Z}$, and some subtyping
relation $\leq$ where $ℕ \leq \mathbb{Z}$. Greed causes us to
instantiate $X$ with ℕ, but in order to type the expression we would
need to instantiate it to $\mathbb{Z}$ instead!

To correct this, we could instead collect these constraints and
solve them only when the function is fully applied to its arguments
(i.e., when we reach a stuck decoration). This mirrors the requirement
in \cite{PT98} that constraints are solved at fully uncurried
applications, maintaining currying but losing a syntactically-obvious
location for synthetic type-argument inference.

We would also need to justify our use of contextual type-argument
inference for checking the types of term arguments. Happily, this does
not appear to be an intractable problem like greed: unlike in
synthesis mode, checking mode for applications in \cite{PT98} does not
require that the synthesized type arguments minimize the result type
of the application, so there is greater freedom in choosing the
instantiations for contextually-inferred type arguments. Hosoya and
Pierce note in \cite{HP99} that the optimal instantiations for these
type arguments are ones that ``maximize the expected type
corresponding to the [argument],'' as the type that the programmer
meant for the argument (if type correct) will be a subtype of
this. Though the informal approach they proposed (and later dismissed)
for inferring the types of hard-to-synthesize terms differs from ours
in the use of a ``slightly ad-hoc'' analysis of arguments, it
anticipated contextual type-argument inference and suggests the way
forward for extending contextual type-argument inference to
subtyping.

\subsection{Bidirectional Type Inference and System F}

\paragraph{\textbf{Predicative Polymorphism}}
The popularity of bidirectional type inference extends well beyond
local inference methods. Dunfield and Krishnaswami\cite{DK13}
introduced a relatively simple and elegant type inference system for predicative
System F using a dedicated application judgment that instantiates type
arguments at term applications. Their application judgment was the
direct inspiration of our own, though there are significant
differences between the two. First, our rules distinguish between
checking the argument of an application with a fully known expected
type and synthesizing its argument when incomplete information is
available to keep meta-variables \textit{spine-local}, whereas
in their approach meta-variables and typing constraints are passed
downwards to check term arguments. Our system also contains the
additional judgment form $\meta$ that theirs does not, again to
contain meta-variables within an application spine.

Approaches to type inference for System F (impredicative and
predicative alike) often make use of some form of subsumption rule to
decrease the required type annotations in terms. A popular basis for
such rules is the ``more polymorphic than'' subtyping relation of
introduced by Odersky and L\"{a}ufer in \cite{OL96} which stratifies
polymorphic and monomorphic types and is able to perform deeply nested
monomorphic type instantiation. This line of work includes \cite{DK13}
above as well as an earlier work by Peyton Jones
et. al. \cite{PVWS07}, both of which are able to infer arbitrary-rank
types in the setting of predicative System F. In contrast our type
inference algorithm supports more powerful impredicative polymorphism
at the cost of significant increase in required type annotations.

\paragraph{\textbf{Impredicative Polymorphism}}
The ``more-polymorphic-than'' subtyping relation for impredicative
System F is undecidable\cite{TU96}, so type inference systems wishing
to use a subsumption rule in this setting must make some
compromises. With \textbf{ML}$^{\textbf{F}}$ \cite{BR03} Le Botlan and
R\'{e}my develop a type language with bounded type quantification and
an inference system using type \textit{instantiation} (a covariant
restriction of subtyping). \textit{Boxy type inference}\cite{VWP06} by
Vytiniotis et al.  uses an idea similar to \cite{OZZ01} of propagating
partial type information (though with a very different implementation)
to allow inference for polymorphic types only in checking mode; its
later development in \textbf{FPH}\cite{VPW08} both simplifies the
specification for type inference and extends boxy types to synthesis
mode to allow ``boxy monotypes'' to be inferred for polymorphic
functions. These inference systems add additional constructs
(resp. bounded quantifications and boxy types) to System F types in
their specification, whereas we reserve our new constructs (decorated
types and prototypes) for the algorithmic rules only. Our use of
first-order matching when typing applications of and arguments to
polymorphic functions can be viewed as a crude form of subtyping via
shallow type instantiation -- it is significantly easier for
programmers to understand but at the same time significantly less
powerful than the subtyping used in the type inference systems above.
    
\begin{acks}
We thank Larry Diehl, Anthony Cantor, and Ernesto Copello for their
feedback on earlier versions of this paper which helped us clarify
some points of terminology and improve the readability of the more
technical sections of the paper.  We gratefully acknowledge NSF
support under award 1524519, and DoD support under award
FA9550-16-1-0082 (MURI program).
\end{acks}